\documentclass[conference, twocolumn]{IEEEtran}

\usepackage{presets}
\usepackage{cite}
\usepackage{amsmath,amssymb,amsfonts}
\usepackage{algorithmic}
\usepackage{graphicx}
\usepackage{textcomp}
\usepackage{xcolor}
\usepackage[left=0.9in,top=0.9in,right=0.9in,bottom=0.9in]{geometry} 
\newcommand{\leqt}{\leq_t}
\newcommand{\lt}{<_t}
\newcommand{\geqt}{\geq_t}
\newcommand{\gt}{>_t}
\newcommand{\remove}[1]{}
\allowdisplaybreaks

\begin{document}
\title{Small-Error Cascaded Group Testing}

\author{\IEEEauthorblockN{Daniel McMorrow}
\IEEEauthorblockA{National University of Singapore \\
mcmorrow@nus.edu.sg}
\and
\IEEEauthorblockN{Nikhil Karamchandani}
\IEEEauthorblockA{IIT Bombay \\
nikhil@ee.iitb.ac.in}
\and
\IEEEauthorblockN{Sidharth Jaggi}
\IEEEauthorblockA{University of Bristol \\
sid.jaggi@bristol.ac.uk}
}

\maketitle

\begin{abstract}
      Group testing concerns itself with the accurate recovery of a set of ``defective'' items from a larger population via a series of tests. While most works in this area have considered the \emph{classical} group testing model, where tests are binary and indicate the presence of at least one defective item in the test, we study the \emph{cascaded} group testing model. In cascaded group testing, tests admit an ordering, and test outcomes indicate the first defective item in the test under this ordering. Under this model, we establish various achievability bounds for several different recovery criteria using both non-adaptive and adaptive test designs when assuming both unconstrained and constrained test sizes. In the constrained test size setting, we also provide a lower bound showing our achievability result is optimal up to logarithmic factors.
\end{abstract}

\section{Introduction}
The goal of group testing is the recovery of a set of ``defective'' items $\cK$ of size $\lvert \cK \rvert = k$ from a population $\cN \defeq \{1,\dots, n\}$, using as few ``tests'' as possible. First considered by Dorfman \cite{dorfman1943detection} as a means of more efficiently testing soldiers for syphilis during World War 2, it has since seen a diverse range of applications such as in DNA testing \cite{gille1991pooling, erlich2015biological, curnow1998pooling}, multiple access channels \cite{wolf1985born, komlos1985asymptotically}, data compression \cite{hong2004group}, random access protocols \cite{inan2017sparse}, and database systems \cite{cormode2005s}. In the canonical model of group testing known as \emph{classical} group testing, a test $t \in [T] \defeq \{1,\dots, T\}$ is specified by an (unordered) subset of items $\cI_t$ with the corresponding test outcome $Y_t \in \{0,1\}$ indicating whether at least one item in the test is defective. That is, $Y_t = 1$ if and only if $\cI_t \cap \cK \neq \emptyset$. In this paper, we study the recently proposed \emph{cascaded} group testing model \cite{mirza2024cascaded}, where each $\cI_t$ admits an ordering, and test outcomes $Y_t \in \cI_t \cup \{0\}$ correspond to the first defective item in $\cI_t$ under this ordering (with an output of $0$ if there are no defectives in the test). Before further discussion on this model however, we survey some of the known results on classical group testing to provide a benchmark for our results. Several versions of the classical group testing model have been considered in the literature. For instance, there are several different forms of recovery criteria that have been studied, with some of the main ones listed below:
\begin{itemize}
    \item In \emph{zero-error} group testing, the goal is to exactly recover the defective set for any possible choice of $\cK$ with probability $1$. That is, we require the estimated defective set $\hat{\cK}$ to satisfy $\bP(\hat{\cK} \neq \cK) = 0$ for given $n, k$ and $\cK$.
    \item In \emph{small-error} group testing with \emph{exact recovery}, the recovery guarantee is weakened to only require $\bP(\hat{\cK} \neq \cK) \to 0$ as $n \to \infty$, where the randomness is over $\cK$ and any potential randomness in the test design and decoder.
    \item In \emph{small-error} group testing with \emph{approximate recovery}, the recovery requirement is further weakened to allow a small number of false positive and/or false negative misclassifications in $\hat{\cK}$.
\end{itemize}
Moreover, the contrasting notions of \emph{adaptive} and \emph{non-adaptive} testing have been considered. In the former, previously observed test outcomes can be used to guide the design of future tests, whereas the latter requires all tests be designed in advance. An argument based on the pigeonhole principle shows that we require $2^T \geq  {n \choose k}$ for exact recovery to be possible in the zero-error setting, implying that $T \geq  \log_2{n \choose k}$ tests are necessary for accurate recovery regardless of the decoder and test design, adaptive or otherwise. A stronger converse bound for the zero-error non-adaptive setting was later provided in 
\cite{d1982bounds, furedi1996onr}, which showed that any non-adaptive group testing strategy requires at least $\min\{n, \Omega\left(k^2\log_k n\right)\}$ tests.
For the small-error setting, a similar lower bound of $T \geq (1 -\epsilon)\log_2 {n \choose k}$ to ensure $\bP(\hat{\cK} \neq \cK) \leq \epsilon$ was given in \cite{chan2011non}, which was subsequently strengthened by Baldassini et al. \cite{baldassini2013capacity}, who showed that $T \geq (1-o(1))\log_2{n \choose k}$ tests are necessary for any scheme to have $\bP(\hat{\cK} \neq \cK) \nrightarrow 1$ as $n \to \infty$. A stronger converse bound specific to non-adaptive test designs has also been derived in \cite{coja2020optimal} for the small-error regime when the number of defectives is assumed to satisfy $k = \Theta(n^\theta)$ for some $\theta \in (0,1)$.

On the achievability side, an adaptive generalized binary splitting algorithm due to Hwang \cite{hwang1972method} achieves near-optimal performance in the zero-error setting (and hence the small-error setting, since their respective lower bounds match up to $o(1)$ factors), requiring $T = \log_2{n \choose k} + k$ tests for successful exact recovery. For non-adaptive testing, $T = \min\{n, \cO(k^2 \log n)\}$ tests have been shown to be sufficient in the zero-error setting, with the $\cO(k^2 \log n)$ term coming from \cite{porat2008explicit} which proposes explicit testing strategies based on coding-theoretic ideas. For the small-error regime, the achievability results depend on the scaling of $k$ relative to $n$. If $k = \Theta(n)$ (i.e., $k$ is \emph{linear} in $n$), then individual testing has been shown to be optimal \cite{aldridge2018individual, bay2022optimal}, whereas if $k = \Theta(n^\theta)$ for $\theta \in [0,1)$ (i.e., $k$ is \emph{sub-linear} in $n$), $T = \cO(k\log n)$ tests have been shown to suffice for exact recovery \cite{chan2014non, aldridge2014group, scarlett2016phase}, which is order optimal since $\log_2 {n \choose k} = \Omega(k \log n)$ under this scaling of $k$. Optimality has also been established up to sharp constant factors in this setting via achievability results in \cite{coja2020information, coja2020optimal}, matching the aforementioned converse for non-adaptive designs. For a comprehensive survey on classical group testing, see \cite{aldridge2019group}.


In this work, we investigate the \textit{cascaded group testing} framework that was recently proposed in \cite{mirza2024cascaded}, which produces
non-binary, position-based feedback. A test in this model is specified by an
ordered subset of items $\cI = \{i_{1}, i_{2}, \ldots, i_{\ell}\}$, and the
output reveals the index of the first defective item encountered in
that order. Note that such a test yields more information than the
classical binary OR test, which only indicates whether any defective
item is present. There are several applications where such cascade feedback is naturally available. For example, consider  network tomography, where probe
packets traverse prescribed routes through a communication network.
A probe either completes the entire route---indicating that every
link on the path is uncongested---or it terminates early and reports
the identity of the first congested link it encounters. Similar feedback is also received in the case of web search, where a search engine, in response to a query,   provides a list of pages which the user sequentially parses  and clicks on the first link it finds relevant to the query. Feedback of this nature has also been extensively studied in the online learning literature under the moniker
\emph{cascading bandits} \cite{kveton2015cascading, kveton2015combinatorial, gan2020cost}, which have been used to model various applications
including ad placement
with ordered user attention and ranked-item selection in online
marketplaces, all of which share the key feature that observations
terminate upon the first salient event in the sequence.
\remove{
In this work, we consider the problem of \emph{cascaded} group testing, which has seen considerably less attention in the literature. In cascaded group testing, tests admit an \emph{ordering}, which represent their ``position'' in the test:
\begin{definition} \label{def:test_order}
    Fix a test $t \in [T]$, and let $\cI_t = \{v_1, \dots, v_{\lvert t \rvert}\}$ be the set of items included in test $t$, where $\lvert t\rvert$ is the number of items included in test $t$. We say that \emph{$i$ appears before $j$ in $t$} (or equivalently $i \lt j$) if there exists $n_i, n_j > 0$ such that $i = v_{n_i}, \ j = v_{n_j}$ and $n_i < n_j$. We say that $i \leqt j$ if $i \lt j$ or $i=j$.
\end{definition}
\noindent A concrete example of this ordering is given as follows:
\begin{example} \label{ex:test_order}
    Fix a test $t \in [T]$ and suppose that $\cI_t = \{3, 1, 2\}$. Then it holds that $3 \lt 1\lt 2$.
\end{example}
\noindent In cascaded group testing, test outcomes indicate the first defective item in the test under its ordering (with $0$ being returned if there are none). That is, for a test $t \in [T]$, the test outcome $Y_t$ is given by
\begin{equation}
    \label{eq:test_outcome} Y_t = \min \{v \in \cI_t : v \in \cK\}
\end{equation}
if $\cI_t \cap \cK \neq \emptyset$, where the minimum is with respect to the ordering induced by $t$, and $Y_t = 0$ otherwise.
\begin{example}
    \label{ex:test_outcome} Under the setup of Example \ref{ex:test_order}, suppose that $\cK = \{2,3\}$. Then, the outcome of test $t$ is given by $Y_t = 3$.
\end{example}
}

It was shown in \cite{mirza2024cascaded} that $\Omega(k^2)$ tests are necessary, and that $(\log \log n)^{2^{\cO(k)}}$ tests suffice for non-adaptive zero-error recovery under cascaded group testing. Note that this can be much smaller than the $\cO(k^2\log n)$ tests shown to be sufficient for classical binary group testing \cite{porat2008explicit} for small $k$, thus demonstrating the significant benefits of the additional information provided by  cascaded testing. In contrast to \cite{mirza2024cascaded}, we consider the small-error setting and both exact as well as approximate recovery under the cascaded group testing model. Next, we highlight the specific contributions made by this work.

\section{Contributions}
\begin{itemize}
    \item We derive new achievability bounds for non-adaptive exact recovery, non-adaptive approximate recovery, and multi-stage adaptive exact recovery, covering both unrestricted test sizes and test sizes upper bounded by some $G \leq n$.
    \item When the size of the defective set $k$ is unknown, we provide an $(\epsilon k, \delta)$-PAC estimator of $k$ using at most $T = \cO(\log(1/\delta)/\epsilon^2)$ tests.
    \item For a uniformly random defective set and small maximum test size $G$, we prove a lower bound showing the multiplicative penalty in our test-size-constrained bounds is unavoidable.
\end{itemize}

\section{Problem Setup}
Given a set $\cN \defeq \{1,\dots, n\}$ of items, there exists a set of defective items $\cK \subseteq \cN$ of size $\lvert \cK \rvert = k$ for some known $k < n$ (except in the case where we are estimating $k$). We assume an arbitrary distribution over the ${n \choose k}$ size-$k$ subsets of $\cN$. In the cascaded group testing model, each test $t \in [T] \defeq \{1, \dots, T\}$ admits an \emph{ordering}, which represents the position of each item in the test. Specifically, for a test $t \in [T]$, items in the test are represented by the ordered set $\cI_t = \{i_1, \dots, i_{\lvert \cI_t \rvert}\} \subseteq \cN$, where the ordering $\leqt$ on $\cI_t$ is defined by $i_r \leqt i_s$ for $i_r, i_s \in \cI_t$, if and only if $r \leq s$ (with $\lt, \ \geqt, \ \gt$ being defined analogously). Note that due to this ordering, we consider sets with the same items, but placed in a different order, as being distinct. 
A concrete example of this ordering is given as follows:
\begin{example} \label{ex:test_order}
    Fix a test $t \in [T]$ and suppose that $\cI_t = \{3, 1, 2\}$. Then it holds that $3 \lt 1\lt 2$.
\end{example}
In this model, each test outcome indicates the first defective item in the test under its respective ordering, with an output of $0$ if there are none. That is, for a test $t \in [T]$, the test outcome $Y_t \in \cI_t \cup \{0\}$ is given by 
\begin{equation}
    \label{eq:test_outcome} Y_t = \underset{\leqt}{\min} \{i \in \cI_t : i \in \cK\},
\end{equation}
\noindent where the notation $\min_{\leqt}$ indicates the minimum is with respect to the ordering $\leqt$ if $\cI_t \cap \cK \neq \emptyset$, and $Y_t = 0$ otherwise. We call the collection of sets $\cI_1,\dots,\cI_T$ alongside their respective orderings a \emph{test design}.

Given the sets $\boldsymbol{\cI} =  (\cI_1, \dots, \cI_T) \in \mathscr{P}(\cN)^T$ and test outcomes $\bfY = (Y_1, \dots, Y_T) \in \bigtimes_{t=1}^T (\cI_t \cup \{0\})$, a (possibly randomized) map
\begin{equation}
    \label{eq:decoder_def} \hat{\cK} :\mathscr{P}(\cN)^{T} \times \left(\bigtimes_{t=1}^T  \big(\cI_t \cup \{0\}\big)\right) \to \mathscr{P}(\cN)
\end{equation}
known as a \emph{decoder} forms an estimate of $\cK$, where $\mathscr{P}(\cN)$ is the power set of $\cN$. Given this estimate, we have the following achievability definitions, depending on the recovery criteria in consideration:

\noindent {\bf Achievability for Exact Recovery.} 
For a given number of tests $T$ and error tolerance $\delta \in (0,1)$, we say that a decoding strategy is \emph{achievable} under the exact recovery criterion if the resulting estimate $\hat{\cK} = \hat{\cK}(\boldsymbol{\cI},\bfY)$ satisfies $\bP(\hat{\cK} \neq \cK) \leq \delta$.


\noindent {\bf Achievability for Subset Recovery.}
Under the same notation and for a parameter $\alpha \in (0,1)$, we define \emph{subset recovery} as a special case of approximate recovery, where the goal is to recover a subset of $\cK$ of size at least $(1-\alpha)k$. Accordingly, we say that a decoding strategy is \emph{achievable} under the subset recovery criterion if $\bP(\hat{\cK} \nsubseteq \cK \cup \lvert \hat{\cK}\rvert < (1-\alpha)k) \leq \delta$. 

In both cases, the probability is over the randomness in $\cK$ and any potential randomness in the test design and decoder. 

\begin{remark}[Prior Distribution of $\cK$] \label{rem:prior_distribution}
     In contrast to the typical assumption of a \emph{uniform prior distribution} over the $n \choose k$ possible defective sets in the classical group testing literature, we impose the much weaker assumption of an \emph{arbitrary prior distribution}. While our upper bounds hold for any such distribution, they become very loose for highly skewed priors. For instance, for a prior distribution placing positive mass on only two possible defective sets, the correct set can be inferred with only one test. We leave a systematic treatment of prior-dependent achievability bounds for future work.
\end{remark}


\section{Main Results}
We now formally state our main results;\footnote{For notational convenience, we assume throughout the theorem statements and proofs that all quantities involving $T$ are integers. If not, the analysis remains unchanged when rounding the respective values of $T$ up to the nearest integer.} the first subsection concerns results without any constraints on the test size, while their constrained test size counterparts are given in the second.
\subsection{Results for Unrestricted Test Sizes}
\begin{theorem}[Non-Adaptive Exact Recovery] \label{thm:non_adaptive_exact} 
    Suppose $\smash{\lvert \cK \rvert = k}$, and let $T = k \log(k/\delta)$ for some $\delta \in (0,1)$. Then, there exists a non-adaptive test design and a decoder such that $\bP(\hat{\cK} \neq \cK) \leq \delta$.
\end{theorem}

\begin{theorem}[Non-Adaptive Subset Recovery] \label{thm:non_adaptive_approximate}
    Suppose $\lvert \cK \rvert = k$, and let $T = 2k \log(1/\alpha) + \frac{\log(1/\delta)}{\alpha}$ for some $\alpha \in (0,1/2]$ and $\delta \in (0,1)$. Then, there exists a non-adaptive test design and a decoder that satisfies $\bP(\hat{\cK} \nsubseteq \cK \cup \lvert \hat{\cK}\rvert < (1-\alpha)k) \leq \delta$.
\end{theorem}

\noindent We require the following definitions for the description of the next theorem:

\begin{definition} \label{def:f_compose}
    For a set $\cA \subseteq \bR$, function $f : \cA \to \bR$, and positive integer $m \in \bZ$, the $m$-fold composition of $f$, denoted by $f^{(m)}$, is defined as
    \begin{equation}
        f^{(m)} = \underbrace{f \circ f \circ \cdots \circ f}_{m \ {\rm times}}.
    \end{equation}
\end{definition}

\begin{definition} \label{def:log_star}
    The function $\log^*: (0,\infty) \to \bZ$ is defined \emph{\cite[Section 3.2]{cormen2022introduction}} as 
    \begin{equation}
        \label{eq:log_star} \log^*(x) = \min\{i \ge 0: \log^{(i)}(x) \leq 1\}.
    \end{equation}
    That is, $\log^*(x)$ is the smallest nonnegative integer $i$ such that $\log^{(i)}(x) \leq 1$, where we define $\log^{(0)}(x) = x$ for consistency.
\end{definition}

\begin{theorem}[Adaptive Exact Recovery with a Limited Number of Stages] \label{thm:mstage_adaptive_exact}
    Suppose that $\lvert \cK \rvert =k$, and let $T = \cO(k \log^{(m)}(k) + k \log(m/\delta))$, where $\delta \in (0,1)$ and $2 \leq m \leq \log^*(k) -1$. Then there exists an $m$-stage adaptive test design and decoder such that $\bP(\hat{\cK} \neq \cK) \leq \delta.$
\end{theorem}

\begin{remark}[Lower Bounds]
    Since each test reveals at most one item, direct lower bounds of $T = k$ and $T = (1-\alpha)k$ apply for exact recovery and subset recovery respectively. For exact recovery, we match this up to a $\log k $ factor in the non-adaptive setting, and a $\log^{(m)} (k)$ factor in the $m$-stage adaptive setting. While we were unable to obtain tighter lower bounds, we conjecture that this $\log k$ factor is unavoidable in the non-adaptive setting (at least in the case that the prior of $\cK$ is uniform). For approximate recovery, our upper bounds are order-optimal if $\alpha = \Theta(1)$, but an interesting future direction of research could be to investigate whether the constant factors can be improved.
\end{remark}

\begin{theorem}[Estimating $k$] \label{thm:k_estimate}
    Suppose that $\lvert \cK \rvert = k$, and that $k$ is unknown. Then for any $\epsilon, \delta \in (0,1)$, there exists a non-adaptive estimator $\hat{K}$ of $k$ such that $\bP( \lvert \hat{K} - k \rvert > \epsilon k) \leq \delta$ using $T = \cO(\log(1/\delta)/\epsilon^2 )$. That is, $\hat{K}$ is an $(\epsilon k, \delta)$-PAC estimator of $k$ using $T = \cO(\log(1/\delta)/\epsilon^2)$ tests. 
\end{theorem}

Comparing these results to their classical group testing counterparts,\footnote{Throughout the upcoming discussion, we implicitly assume  $k = \Theta(n^\theta)$ for some $\theta \in [0,1)$, since this is the range of $k$ mainly considered in the classical group testing literature.} we observe the following:

\begin{itemize}
    \item When $\delta = \Omega({\rm poly}(1/k))$, Theorem \ref{thm:non_adaptive_exact} implies that $T = \cO(k \log k)$ tests suffice for exact recovery, which may be much smaller than the $\Omega(k \log n)$ tests required for exact recovery in group testing for certain values of $k$ (e.g., $k = \Theta(\log n)$), while still ensuring vanishing error probability as $n \to \infty$. 
    \item If $\delta = \Omega({\rm poly}(1/k))$ and $\alpha$ is constant, Theorem \ref{thm:non_adaptive_approximate} implies that $\cO(k)$ tests suffice for successful recovery, which is much smaller than the $\Omega(k \log n)$ tests needed for approximate recovery in the classical setting \cite{scarlett2016phase}.
    \item For the multi-stage adaptive setting, if $m = \cO(1)$ and $\delta = \Omega(1)$, Theorem \ref{thm:mstage_adaptive_exact} states that $T = \cO(k \log^{(m)}(k))$ tests suffice for exact recovery, which is much smaller than the $\Omega(k \log n)$ tests required in the analogous classical setting. 
    \item If $\epsilon$ and $\delta$ are both constant, then Theorem \ref{thm:k_estimate} implies $T = \cO(1)$ tests, suffice to form an $(\epsilon k, \delta)$-PAC estimator of $k$. This is significantly smaller than the $\Omega( \log \log k)$ tests required under classical testing when designing tests adaptively \cite{falahatgar2016estimating}, or the $\Omega(\log (n) /\log^{(j)}(n))$ tests required when designing tests non-adaptively for sufficiently large $n$, where $j$ is an arbitrary integer \cite{bshouty2025improved}.
\end{itemize}

The above comparison demonstrates that the extra information provided from cascaded tests can significantly reduce the number of tests needed for accurate recovery. Another point of interest is that contrary to classical group testing, the number of tests required for successful recovery using cascaded tests is independent of the population size $n$.

We now present analogous results under the additional assumption that test sizes are bounded.
\subsection{Results for Bounded Test Sizes}
\begin{theorem}[Non-Adaptive Exact Recovery] \label{thm:non_adaptive_exact_leq_G}
    Consider the same setup as in Theorem \ref{thm:non_adaptive_exact}, but with the extra constraint that each test can only include at most $G$ items. Then, if $T = \frac{k}{1-\exp(-kG/n)} \log(k/\delta)$, there exists a non-adaptive test design and decoder such that $\bP(\hat{\cK} \neq \cK) \leq \delta$.   
\end{theorem}
\begin{theorem}[Non-Adaptive Subset Recovery] \label{thm:non_adaptive_approximate_leq_G}
    Consider the same setup as in Theorem \ref{thm:non_adaptive_approximate}, but with the extra constraint that each test can only include at most $G$ items. Then, if $T = \frac{1}{1-\exp(-kG/n)}\big(2k \log(1/\alpha) + \frac{\log(1/\delta)}{\alpha}\big)$, there exists a non-adaptive test design and decoder such that $\bP(\hat{\cK} \nsubseteq \cK \cup \lvert \hat{\cK}\rvert < (1-\alpha)k) \leq \delta$.
\end{theorem}
\begin{theorem}[Adaptive Exact Recovery with a Limited Number of Stages] \label{thm:mstage_adaptive_exact_leq_G}
    Consider the same setup as in Theorem \ref{thm:mstage_adaptive_exact}, but with the extra constraint that each test can only include at most $G$ items. Then, if $T = \cO\big(\frac{1}{1-\exp(-kG/n)}(k \log^{(m)}(k) + k \log(m/\delta))\big)$, there exists an $m$-stage adaptive test design and decoder such that $\bP(\hat{\cK} \neq \cK) \leq \delta$.
\end{theorem}

\begin{remark}[Discussion on Test Sizes] \label{rem:limited_test_size}
    Theorems \ref{thm:non_adaptive_exact_leq_G}-\ref{thm:mstage_adaptive_exact_leq_G} show that constraining tests to be of size at most $G$ imposes a multiplicative ``test-size penalty'' of $\frac{1}{1 - \exp(-kG/n)}$ on the number of tests required, compared to Theorem \ref{thm:non_adaptive_exact}. When $G = \Omega(n/k)$, this factor scales as $\Theta(1)$, whereas for $G = o(n/k)$, it scales as $\Theta(n/(Gk))$. This implies using tests of size $G = \Omega(n/k)$ suffice to match the order of $T$ in Theorems \ref{thm:non_adaptive_exact}-\ref{thm:mstage_adaptive_exact}, while $T = \cO((n/G)\log k) = \omega(k \log k)$ tests suffice if $G = o(n/k)$.
\end{remark}
In the case that $G = \cO(n/k)$ and the defective set has a uniform prior, this $\frac{n}{G}$ factor is unavoidable for any non-adaptive test design and decoder, as demonstrated in the following theorem:
\begin{theorem} \label{thm:small_G_lower}
    Suppose that $\cK$ is uniformly distributed over all ${n \choose k}$ possibilities, and assume that $G \leq \frac{n-k}{k}$. Then for any non-adaptive test design $\boldsymbol{\cI}$, decoder $\hat{\cK}$ and $\delta \in [\frac{1}{2(n-k)},1)$,
    \begin{equation}
        \label{eq:small_G_lower} T \geq (1-2\delta)\frac{n-k}{G}
    \end{equation}
    tests are required to ensure that $\bP(\hat{\cK} \neq \cK) \leq \delta$. 
\end{theorem}
Similar to the unconstrained test size setting, our upper bound differs from our lower bound by a factor of $\log k$. We leave the problem of establishing order-wise matching upper/lower bounds to future work.

\section{Proofs}
Here we provide the proofs of Theorems \ref{thm:non_adaptive_exact}-\ref{thm:small_G_lower}. The first subsection contains the proofs of Theorems \ref{thm:non_adaptive_exact}-\ref{thm:k_estimate} (i.e., the results where any test size is allowed), while the second subsection contains the proofs of Theorems \ref{thm:non_adaptive_exact_leq_G}-\ref{thm:small_G_lower} (i.e., the results with bounded test sizes).

\medskip
\noindent {\bf \underline{Proof of Theorems \ref{thm:non_adaptive_exact}-\ref{thm:k_estimate}}}
\medskip

\noindent When there are no constraints on the test size, we design our tests by letting $\cI_t \sim {\rm Unif}(\Sigma_\cN)$ independently for all $t$, where $\Sigma_\cN$ is the set of all permutations of $\cN$. That is, each test contains all items in $\cN$ and the ordering $\leqt$ is given by a uniformly random permutation of $\cN$. Throughout, we form our estimate $\hat{\cK}$ by taking it to be the union of test outcomes (i.e., $\hat{\cK} = \cup_{t=1}^T \{Y_t\}$).

\subsection{Proof of Theorem \ref{thm:non_adaptive_exact} (Non-Adaptive Exact Recovery)}
Forming $\hat{\cK}$ in the manner described above leads to a coupon collector's problem \cite{ferrante2014coupon}, where the decoder outputs all of the defective items ``collected'' by the tests. Thus, to have $\hat{\cK} = \cK$ we require that every item in $\cK$ appears as the first defective in at least one test (or equivalently, that every defective item is ``collected''). The probability of a defective item $i \in \cK$ being the first defective in a test (and thus correctly recovered) is $\frac{1}{k}$, since each permutation is sampled uniformly at random. Thus, the probability of $i \in \cK$ \emph{not} being the first defective item in \emph{any} test is $(1 - 1/k)^T$ due to the independence of tests. Taking a union bound over the $k$ items in $\cK$ yields
\begin{equation}
    \bP(\hat{\cK} \neq \cK) \leq k\left(1 - \frac{1}{k}\right)^T \leq k \exp\left(-\frac{T}{k}\right).
\end{equation}
Choosing $T = k \log(k/\delta)$ ensures that $\bP(\hat{\cK} \neq \cK) \leq \delta$, as claimed.

\subsection{Proof of Theorem \ref{thm:non_adaptive_approximate} (Non-Adaptive Subset Recovery)}
Forming $\hat{\cK}$ in the manner described above once again leads to a (partial) coupon collector's problem. For $\alpha \in \left (0,\frac{1}{2} \right ]$, an error therefore occurs if at least one subset of $\cV \subseteq \cK$ of size $\lvert \cV \rvert = \alpha k$ has no items appearing as the first defective in any test, since this implies that less than $(1-\alpha)k$ distinct defectives are recovered. Accordingly, defining the error event $\cE \defeq \{\hat{\cK} \nsubseteq \cK \cup \lvert \hat{\cK}\rvert < (1-\alpha)k\}$, the error probability satisfies
\begin{align}
    \label{eq:approximate_error_prob_1} \bP(\cE) &= \bP\left(\bigcup_{\cV \subseteq \cK : \lvert \cV \rvert = \alpha k} \ \bigcap_{t=1}^T \, \{Y_t \notin \cV\}\right) \\
    \label{eq:approximate_error_prob_2} &\leq \sum_{\cV \subseteq \cK : \lvert \cV \rvert = \alpha k} \ \prod_{t=1}^T\bP(Y_t \notin \cV) \\
    \label{eq:approximate_error_prob_3} &= {k \choose \alpha k}(1-\alpha)^T,
\end{align}
where \eqref{eq:approximate_error_prob_2} follows from a union bound and the independence of test outcomes, and \eqref{eq:approximate_error_prob_3} follows since $\bP(Y_t \notin \cV) = 1-\alpha$ for all $\cV$ and $t$ due to the symmetry of the test design. Using the bounds $(1-\alpha)^T \leq e^{-\alpha T}$ and ${k \choose \alpha k} \leq 2^{k H_2(\alpha)}$, where $H_2(\alpha)$ is the binary entropy function measured in bits, we can further bound the error by
\begin{align}
    \label{eq:approximate_error_prob_4} \bP(\cE) &\leq \exp\left(k \log (2)H_2(\alpha) - \alpha T\right) \\
    \label{eq:approximate_error_prob_5} &\leq \exp\left(2\alpha k \log\Big(\frac{1}{\alpha}\Big) - \alpha T\right),
\end{align}
where \eqref{eq:approximate_error_prob_5} uses the bound $H_2(\alpha) \leq 2\alpha \log_2(1/\alpha)$ for $\alpha \in (0, \frac{1}{2}]$.\footnote{This can easily be seen by the fact that the first term in the binary entropy function dominates for $\alpha \in (0,\frac{1}{2}]$.} Finally, choosing $T = 2k\log(1/\alpha) + \frac{\log(1/\delta)}{\alpha}$ ensures that $\bP(\cE) \leq \delta$.

\subsection{Proof of Theorem \ref{thm:mstage_adaptive_exact} (Adaptive Exact Recovery with a Limited Number of Stages)}
To form our multi-stage adaptive exact recovery algorithm, we combine the algorithms for non-adaptive exact recovery and subset recovery. Namely, for subset recovery parameters $\alpha_1, \dots, \alpha_{m-1} \in (0,1)$ we do the following:

\begin{itemize}
    \item In the first stage, run the non-adaptive subset recovery algorithm from Theorem \ref{thm:non_adaptive_approximate} to recover a $1-\alpha_1$ fraction of $\cK$.
    \item In stages $\ell= 2,3,\dots,m-1$, run the same non-adaptive subset recovery algorithm from Theorem \ref{thm:non_adaptive_approximate} to recover a $1- \alpha_j$ fraction of the remaining $k\prod_{j=1}^{\ell-1}\alpha_j$ items.
    \item In stage $m$, run the non-adaptive exact recovery algorithm from Theorem \ref{thm:non_adaptive_exact} to recover the remaining $k\prod_{j=1}^{m-1}\alpha_j$ items.
\end{itemize}
Assigning failure probability $\delta/m$ to each stage (so the overall error probability is at most $\delta$ by a union bound), Theorem \ref{thm:non_adaptive_approximate} implies that
\begin{equation}
    \label{eq:stage_ell_tests} T_\ell = 2\left(\prod_{j=0}^{\ell - 1} \alpha_j\right) k\log\Big(\frac{1}{\alpha_\ell}\Big) + \frac{\log(m/\delta)}{\alpha_\ell}
\end{equation}
tests suffice for an error probability $\delta/m$ in each stage $\ell = 1, \dots, m-1$, where we define $\alpha_0 =1$ for consistency. For the final stage, Theorem \ref{thm:non_adaptive_exact} implies that
\begin{equation}
    \label{eq:stage_m_tests} T_m = \left( \prod_{j=1}^{m-1}\alpha_j\right) k\log\left(\frac{mk\prod_{j=1}^{m-1}\alpha_j}{\delta}\right)
\end{equation}
tests suffice for an error probability of at most $\delta/m$. Thus, the total number of tests sufficient for $\bP(\hat{\cK} \neq \cK) \leq \delta$ is
\begin{align}
    T &=  \sum_{\ell=1}^{m-1} \left(2\left(\prod_{j=0}^{\ell - 1} \alpha_j\right) k\log\Big(\frac{1}{\alpha_\ell}\Big) + \frac{\log(m/\delta)}{\alpha_\ell}\right) \nonumber \\
     \label{eq:mstage_num_tests_1} &+ \left( \prod_{j=1}^{m-1}\alpha_j\right) k\log\left(\frac{mk\prod_{j=1}^{m-1}\alpha_j}{\delta}\right) \\
     &\leq2 k \log\Big(\frac{1}{\alpha_1}\Big) + \frac{\log(m/\delta)}{\alpha_1} + \alpha_{m-1} k\log\left(\frac{mk}{\delta}\right) \nonumber \\
     \label{eq:mstage_num_tests_2}  &+\sum_{\ell=2}^{m-1} \left(2\alpha_{\ell - 2}\alpha_{\ell -1 }k\log\Big(\frac{1}{\alpha_\ell}\Big)+ \frac{\log(m/\delta)}{\alpha_\ell} \right)
\end{align}
where in \eqref{eq:mstage_num_tests_2} we upper bounded $\alpha_j$ by $1$ except for $\alpha_{\ell -1}, \alpha_{\ell - 2}$ in stages $\ell = 2,\dots, m-1$, and upper bounded all $\alpha_j$ by $1$ except for $\alpha_{m-1}$ in stage $\ell = m$. We then make the choice $\alpha_\ell = 1/\log^{(m-\ell)}(k)$ for $\ell \geq 1$
(see Definition \ref{def:f_compose}). Note that since $m \leq \log^*(k) - 1$, the $\alpha_\ell$ terms are well-defined. Furthermore, the choices of $\alpha_\ell$ satisfy $\log(\frac{1}{\alpha_\ell}) = \log^{(m-\ell +1)}(k)=\frac{1}{\alpha_{\ell - 1}}$, which means that
\begin{align}
    T&= 2k \log^{(m)}(k) + \log^{(m-1)}(k) \cdot \log\Big(\frac{m}{\delta}\Big)  \nonumber \\
    & + \frac{k}{\log(k)}\log\Big(\frac{mk}{\delta}\Big)+ \sum_{\ell = 2}^{m-1}\Bigg(2\alpha_{\ell -2}k  \nonumber \\
    \label{eq:t_multi_1}&\quad \quad \qquad \qquad \qquad \qquad \quad + \log^{(m-\ell)}(k) \cdot \log\Big(\frac{m}{\delta}\Big)\Bigg) \\
    &= \cO\left(k \log^{(m)}(k) + k \log\Big(\frac{m}{\delta}\Big)\right)
\end{align}
tests suffice for $m$-stage adaptive exact recovery.
\subsection{Proof of Theorem \ref{thm:k_estimate} (Estimating $k$)}
We introduce the random variables $\{Z_{t}\}_{t=1}^T$, which denote the position of the first defective item in each test $t$. That is, each $Z_t$ is given by $\smash{\lvert \{i \in \cI_t : i \leqt Y_t\}\rvert}$.
To characterize the distribution of each $Z_t$, we briefly pause to consider the following setting. Suppose there is an urn containing $n$ balls, with $k$ of them being ``good". At each time step, a ball is chosen uniformly at random from the urn without replacement until the first ``good" ball is removed. Letting $X$ denote the number of draws until the first ``good" ball, \cite[Theorems 2.5 and 2.8]{ahlgren2014probability} gives the following results on the mean and variance of $X$:
\begin{align}
    \label{eq:draws_mean} \bE[X] &= \frac{n+1}{k+1} \\
    \label{eq:draws_var} {\rm Var}(X) &= \frac{k(n-k)(n+1)}{(k+2)(k+1)^2} \leq \left(\frac{n+1}{k+1}\right)^2 = \bE[X]^2
\end{align}
Since each test is a uniformly random permutation of $\cN$, they can alternatively be viewed as being designed in the following sequential manner:
\begin{enumerate}
    \item Set $\cI_t = \emptyset$.
    \item Choose an item $I \sim {\rm Unif}(\cN \setminus \cI_t)$.
    \item Update $\cI_t \leftarrow \cI_t \cup \{I\}$.
    \item Repeat (2)-(3) until $\cI_t = \cN$.
\end{enumerate}
This exactly matches the previously considered urn setting, meaning that the random variables $\{Z_t\}_{t=1}^T$ follow the same distribution as $X$. With this in mind, we set our (non-adaptive) estimator $\hat{K}$ of $k$ to be 
\begin{equation}
    \label{eq:khat} \hat{K} = \frac{n+1}{\hat\mu(\{Z_t\}_{t=1}^T)} - 1,
\end{equation}
where $\hat{\mu}: \bR^T \to \bR$ is a robust estimator of $\bE[Z_t]$ to be specified later. To give some intuition behind the choice of $\hat{K}$, note that due to the distribution of each $Z_t$ and \eqref{eq:draws_mean}, it holds that $\bE[Z_t] = \frac{n+1}{k+1}$ and\footnote{Note that the variance, while bounded, is unknown due to its dependence on $k$. Thus, we require a robust estimator that does not require knowledge of the variance, such as the median of means or trimmed mean estimators.} ${\rm Var}(Z_t) = \frac{k(n-k)(n+1)}{(k+2)(k+1)^2}$. This implies that when $\hat{\mu}(\{Z_t\}_{t=1}^T) = \bE[Z_t]$, we have that $\hat{K} = k$, and that values of $\hat{\mu}(\{Z_t\}_{t=1}^T)$ near its mean will lead to $\hat{K}$ being near the true value of $k$. Thus, analyzing the performance of the estimator $\hat{K}$ reduces to a concentration analysis of $\hat{\mu}(\{Z\}_{t=1}^T)$. To formalize this, consider the probability $\bP(\hat{K} > (1+\epsilon)k)$ for some $\epsilon > 0$. Direct algebraic manipulations then imply that 
\begin{align}
    \label{eq:k_to_z_1} \bP(\hat{K} > (1+\epsilon)k) &= \bP\left(\frac{n+1}{\hat{\mu}(\{Z_t\}_{t=1}^T)}-1 > (1+\epsilon)k\right) \\
    \label{eq:k_to_z_2} &= \bP\left(\hat{\mu}(\{Z_t\}_{t=1}^T) < \frac{n+1}{(1+\epsilon)k + 1}\right) \\
    \label{eq:k_to_z_3} &= \bP\left(\hat{\mu}(\{Z_t\}_{t=1}^T) < \frac{1}{1+\frac{\epsilon k}{k+1}} \frac{n+1}{k+1}\right) \\
    \label{eq:k_to_z_4} &\leq \bP\left(\hat{\mu}(\{Z_t\}_{t=1}^T) < \frac{1}{1+\frac{\epsilon}{2}}\frac{n+1}{k+1}\right) \\
    \label{eq:k_to_z_5} &= \bP\left(\hat{\mu}(\{Z_t\}_{t=1}^T) < (1-\gamma)\bE[Z]\right),
\end{align}
where \eqref{eq:k_to_z_4} follows since $k \geq 1$, and \eqref{eq:k_to_z_5} arises from setting $\gamma = \frac{\epsilon}{2+\epsilon}$. Analogous calculations imply that $\bP(\hat{K} < (1-\epsilon)k) \leq \bP(\hat{\mu}(\{Z_t\}_{t=1}^T) < (1-\gamma'')\bE[Z_t])$ for $\epsilon \in(0,1)$, where $\gamma' = \frac{\epsilon}{1-\epsilon}$. A union bound and the fact that $\gamma \leq \gamma'$ when $\epsilon \in (0,1)$ thus jointly imply that
\begin{equation}
    \label{eq:k_error_bound} \bP( \lvert \hat{K} - k \rvert > \epsilon k) \leq \bP \big(\big\lvert \hat{\mu}(\{Z_t\}_{t=1}^T) - \bE[Z_t] \big\rvert > \gamma \bE[Z_t]\big).
\end{equation}
It now remains to bound the right hand side of \eqref{eq:k_error_bound}, which can be done using an off-the-shelf robust mean estimator. We use the median of means estimator (see for instance \cite[Theorem 2]{lugosi2019mean}), which for a sequence of i.i.d. random variables $\{X_i\}_{i=1}^n$ with mean $\mu$ and variance $\sigma^2$, forms an estimate $\hat{\mu}(\{X_i\}_{i=1}^n)$ of $\mu$ such that if $n = MK$, where $K =  8 \log(1/\delta)$, and $M \in \bN$,
\begin{equation}
    \label{eq:catoni} \lvert \hat{\mu}(\{X_i\}_{i=1}^n - \mu \rvert \leq \sqrt{\frac{32\sigma^2}{n}\log\Big(\frac{1}{\delta}\Big)}
\end{equation}
holds with probability at least $1-\delta$. Applying this bound to the right hand side of \eqref{eq:k_error_bound} and using the variance bound in \eqref{eq:draws_var} implies that the following holds with probability at least $1-\delta$:
\begin{align}
    \label{eq:k_error_catoni} & \big\lvert \hat{\mu}(\{Z_t\}_{t=1}^T) - \bE[Z_t] \big\rvert\leq  \sqrt{\frac{32\bE[Z_t]^2}{T}\log\Big(\frac{1}{\delta}\Big)}.
\end{align}
Finally, choosing $T = \frac{32}{\gamma^2}\log(1/\delta)$ (i.e., $M = 4/\gamma^2$) ensures that $\lvert \hat{\mu}(\{Z_t\}_{t=1}^T) - \bE[Z_t]\rvert \leq \gamma \bE[Z_t]$ with probability at least $1-\delta$, which in turn implies $\lvert \hat{K} - k\rvert \leq \epsilon k$ with probability at least $1-\delta$ via \eqref{eq:k_error_bound}. Since $\gamma = \frac{\epsilon}{2+\epsilon} = \Omega(\epsilon)$ for $\epsilon \in (0,1)$, our claim follows.

\medskip 
{\noindent \bf \underline{Proofs of Theorems \ref{thm:non_adaptive_exact_leq_G}-\ref{thm:small_G_lower}}}
\medskip

\noindent When test sizes are constrained to have size at most $G$, we design each test $t$ by first choosing a set $\cG_t \subset \cN$ of size $\lvert \cG_t \rvert = G$ uniformly at random, then setting $\cI_t \sim {\rm Unif}(\Sigma_{\cG_t})$, where we recall $\Sigma_{\cG_t}$ is the set of permutations of $\cG_t$. Once again, we form our estimate $\hat{\cK}$ by taking it to be the union of test outcomes (i.e., $\hat{\cK} = \cup_{t=1}^T \{Y_t\}$).

\subsection{Proof of Theorem \ref{thm:non_adaptive_exact_leq_G} (Non-Adaptive Exact Recovery with Test Size Constraints)}

The idea behind the proof of Theorem \ref{thm:non_adaptive_exact_leq_G} is similar to that of Theorem \ref{thm:non_adaptive_exact}, with some extra technical challenges arising from the smaller test sizes. Namely, we seek to lower bound the probability that a defective item $i \in \cK$ is recovered (or ``collected'') in a \emph{given} test $t$, and then use this to establish an upper bound on the probability that it is not recovered in \emph{any} test. Toward this, observe that the probability that $i$ is recovered (i.e., $\bP(Y_t = i)$) can be written as $\bP(Y_t = i) = \bP( i \in \cG_t \cap Y_t = i)$, since $i$ cannot be recovered if $i \notin \cG_t$. To evaluate this, we decompose the intersection of the events to obtain $\smash{\bP(Y_t = i) = \bP(i \in \cG_t) \bP(Y_t = i \mid i \in \cG_t)}$, and evaluate each term separately. Since $\cG_t$ is chosen uniformly at random, it follows that $\bP(i \in \cG_t) = \frac{G}{n}$. To evaluate $\bP(Y_t = i \mid i \in \cG_t)$, we average over the size of $\cG_t \cap \cK$ to obtain
\begin{align}
    \bP(Y_t = i \mid i \in \cG_t) &= \sum_{s=1}^{\min\{G,k\}} \bP(\lvert \cG_t \cap \cK\rvert = s \mid i \in \cG_t) \nonumber \\
    \label{eq:G_cap_K_average} & \qquad \times \bP(Y_t = i \mid i \in \cG_t, \lvert \cG_t \cap \cK \rvert = s).
\end{align}
We now evaluate each term in the summand, beginning with the first. Conditioned on the event $\{i \in \cG_t\}$, a simple counting argument reveals that the number of sets $\cG_t$ such that $\lvert \cG_t \cap \cK \rvert = s$ is ${k-1 \choose s-1}{n - k \choose G - s}$. Combining this with the fact that $\cG_t$ is chosen uniformly at random, we can conclude that
\begin{equation}
    \label{eq:conditional_size_s} \bP(\lvert \cG_t \cap \cK\rvert = s \mid i \in \cG_t) = \frac{{k-1 \choose s-1}{n - k \choose G - s }}{{n - 1 \choose G -1}}
\end{equation}
for any $1 \leq s \leq \min\{G, k\}$. For the second term, the fact that $\cI_t \sim {\rm Unif}(\Sigma_{\cG_t})$ immediately implies that
\begin{equation}
    \label{eq:conditional_first_item} \bP(Y_t = i \mid i \in \cG_t, \, \lvert \cG_t \cap \cK \rvert = s) = \frac{1}{s}.
\end{equation}
Combining everything, we can thus express $\bP(Y_t = i)$ as
\begin{align}
    \label{eq:P_Yt_i_1} \bP(Y_t = i) &= \frac{G}{n} \sum_{s=1}^{\min\{G,k\}} \frac{{k-1 \choose s-1}{n - k \choose G - s }}{s {n - 1 \choose G -1}} \\
    \label{eq:P_Yt_i_2} &= \frac{1}{{n \choose G}} \sum_{s=1}^{\min\{G,k\}} \frac{{k-1 \choose s-1}{n - k \choose G - s }}{s},
\end{align}
where \eqref{eq:P_Yt_i_2} factors out $\frac{1}{{n-1 \choose G-1}}$ and uses $\frac{n}{G} {n-1\choose G-1} = {n \choose G}$. Using the identity $\frac{{k-1 \choose s -1}}{s} = \frac{{k \choose s}}{k}$ alongside adding and subtracting ${n-k \choose G}$, we can write $\bP(Y_t = i)$ as
\begin{align}
    \bP(Y_t = i) &= \frac{1}{k{n \choose G}} \Bigg(\Bigg[\sum_{s=0}^{\min\{G,k\}} {k \choose s}{n - k \choose G - s}\Bigg] \nonumber \\
    \label{eq:P_Yt_i_3} &\qquad \qquad \qquad\qquad \qquad - {n-k \choose G}\Bigg).
\end{align}
Applying Vandermonde's identity (see e.g., \cite[Example 2.3.4]{chen1992principles}) to the sum, it  follows that
\begin{align}
    \label{eq:P_Yt_i_4} \bP(Y_t =i) &= \frac{1}{k{n \choose G}}\left({n \choose G} - {n - k \choose G}\right) \\
    \label{eq:P_Yt_i_5} &= \frac{1}{k}\left(1 - \frac{{n- k \choose G}}{{n \choose G}}\right) \\
    \label{eq:P_Yt_i_6} &= \frac{1}{k}\left(1 - \frac{(n-k)\cdots (n-k-G+1)}{n\cdots (n-G+1)}\right) \\
    \label{eq:P_Yt_i_7} &\geq \frac{1}{k}\left( 1- \Big(1-\frac{k}{n}\Big)^G\right) \\
    \label{eq:P_Yt_i_8} &\geq \frac{1}{k}(1- e^{-\frac{kG}{n}}),
\end{align}
where \eqref{eq:P_Yt_i_7} upper bounds $\frac{n-k-i}{n-i}$ by $1-\frac{k}{n}$ for each $i = 0,\dots, G-1$, and \eqref{eq:P_Yt_i_8} uses $1 - x \leq e^{-x}$ for $x \geq 0$, and applies this with $x = \frac{k}{n}$. This gives us our final lower bound on $\bP(Y_t = i)$. From here, we can deduce that the probability that any item $i \in \cK$ is \emph{not} recovered is upper bounded by $(1- \frac{1}{k}(1- e^{-\frac{kG}{n}}))^T$, which can be further upper bounded by $\exp(-\frac{T}{k}(1 - e^{-\frac{kG}{n}}))$. Taking a union bound over all $i \in \cK$ thus implies that
\begin{equation}
    \bP(\hat{\cK} \neq \cK) \leq k \exp\left(-\frac{T}{k}\big(1 - e^{-\frac{kG}{n}}\big)\right).
\end{equation}
Choosing $T = \frac{k}{1- \exp(-kG/n)}\log(k/\delta)$ ensures the error probability is at most $\delta$, completing the proof.

\subsection{Proof of Theorem \ref{thm:non_adaptive_approximate_leq_G} (Non-Adaptive Approximate Recovery with Test Size Constraints)}
Since the proof of Theorem \ref{thm:non_adaptive_approximate_leq_G} is very similar to that of Theorem \ref{thm:non_adaptive_approximate}, we only give an outline, and highlight the key differences. Instead of having $\bP(Y_t \notin \cV) = 1-\alpha$, it is now bounded by
\begin{align}
    \label{eq:notin_V_1} \bP(Y_t \notin \cV) &= 1- \bP(Y_t \in \cV) \\
    \label{eq:notin_V_2} &= 1 - \bP\left(\bigcup_{i \in \cV} \{Y_t = i\}\right) \\
    \label{eq:notin_V_3} &= 1 - \sum_{i \in \cV} \bP(Y_t = i) \\
    \label{eq:notin_V_4} &\leq 1 - \alpha\big(1 - e^{-\frac{-kG}{n}}\big),
\end{align}
where \eqref{eq:notin_V_3} follows since the events $\smash{\{Y_t = i\}_{i \in \cV}}$ are mutually disjoint, and \eqref{eq:notin_V_4} uses the lower bound on $\bP(Y_t = i)$ derived in the previous section. Following the same steps as in the proof of Theorem \ref{thm:non_adaptive_approximate}, we can conclude that $T = \frac{1}{1 - \exp(-kG/n)}\big(2k \log(1/\alpha) + \frac{\log(1/\delta)}{\alpha}\big)$ tests suffice to obtain an error probability at most $\delta$, as claimed.
\subsection{Proof of Theorem \ref{thm:mstage_adaptive_exact_leq_G} (Adaptive Exact Recovery with a Limited Number of Stages and Test Size Constraints)}

The proof of Theorem \ref{thm:mstage_adaptive_exact_leq_G} is identical to that of Theorem \ref{thm:mstage_adaptive_exact}, except for the fact that the number of tests $T_\ell$ in each stage is now divided by $1 - \exp(-kG/n)$. This gives the desired result.

\subsection{Proof of Theorem \ref{thm:small_G_lower} (Constrained Test Size Lower Bound)}
The proof of Theorem \ref{thm:small_G_lower} revolves around the concept of ``masking'', which has seen success in proving converse bounds in classical group testing \cite{aldridge2018individual, coja2020optimal, bay2022optimal, mcmorrow2026optimal}. Analogously to the classical setting, we say that an item $i \in \cN$ is \emph{masked} if one of the two following conditions hold: (i) For every $t \in [T]$ such that $i \in \cI_t$, it holds that $i \gt Y_t$; (ii) it is not included in any test. Intuitively, an item being masked means that the test outcomes provide no information about the item's defectivity status, making it difficult to correctly classify it (and hence correctly recovery $\cK$). In more detail, we aim to show that if $T \leq (1-2\delta)\frac{n-k}{G}$, there is at least one masked defective item $i \in \cK$ and one masked non-defective item $j \in \cN \setminus \cK$ with probability at least $2\delta$. Given this, we then aim to show that the MAP decoder has error probability decoder at least $\frac{1}{2}$, which implies the same for \emph{any} decoder.

We begin with showing that the above condition on $T$ implies there is at least one masked non-defective. Note that since each test contains at most $G$ items, it can contain at most $G$ non-defectives in each test. Thus, the total number of non-defectives appearing in the $T$ tests is upper bounded by
\begin{align}
    \label{eq:revealed_nd_ub_1} TG &\leq (1-2\delta)(n-k) \\
    \label{eq:revealed_nd_ub_2} &\leq n - k - 1,
\end{align}
where \eqref{eq:revealed_nd_ub_2} uses $\delta \geq \frac{1}{2(n-k)}$ as assumed in the theorem statement. Thus, at least one non-defective appears in no tests, implying it is masked.

To obtain a lower bound on the number of masked defectives, we begin by lower bounding $\bP(Y_t = 0)$ (i.e., the probability of a negative test), which in turn will give us an upper bound on $\bP(Y_t \neq 0)$. To do so, fix an arbitrary test $t$ of size $\lvert \cI_t \rvert = g_t \leq G$. Given this, the probability (over the randomness in $\cK$) of a negative test then equals
\begin{equation}
    \label{eq:negative_test_prob} \bP(Y_t = 0) = \frac{{n - g_t \choose k}}{{n \choose k}},
\end{equation}
since $\cK$ is uniform over all ${n \choose k}$ possibilities. Expanding out the fraction and noting that $\frac{n- g_t - i}{n - i} \geq \frac{n - g_t - k + 1}{n  - k + 1}$ for all $0 \leq i \leq k-1$, we can therefore lower bound \eqref{eq:negative_test_prob} by
\begin{align}
    \label{eq:negative_test_lower_1} \bP(Y_t = 0 ) &\geq \left(1 - \frac{g_t}{n-k+1}\right)^k \\
    \label{eq:negative_test_lower_2} &\geq \left(1 - \frac{G}{n-k+1}\right)^k \\
    \label{eq:negative_test_lower_3} &\geq 1 - \frac{kG}{n-k+1} \\
    \label{eq:negatice_test_lower_4} &\geq 1 - \frac{kG}{n-k},
\end{align}
where \eqref{eq:negative_test_lower_2} uses $g_t \leq G$ and \eqref{eq:negative_test_lower_3} follows from Bernoulli's inequality and the fact that $\smash{G \leq \frac{n-k}{k}}$ by assumption. Therefore, it follows that $\bP(Y_t \neq 0) \leq \frac{kG}{n-k}$. Now, define
\begin{equation}
    \label{eq:calY} \cY \defeq \bigcup_{t=1}^T \; \{Y_t\}
\end{equation}
as the set of defective items recovered by the $T$ tests. Since the same defective item may be the outcome of multiple tests, the size of $\cY$ is trivially upper bounded by 
\begin{equation}
    \label{eq:abs_calY_bound} \lvert \cY \rvert \leq \sum_{t=1}^T \mathds{1}\{Y_t \neq 0\}
\end{equation}
(i.e., the number of positive tests). Taking expectations on both sides and using the upper bound on $\bP(Y_t \neq 0)$, it follows that
\begin{equation}
    \label{eq:mean_abs_calY_bound} \bE\lvert \cY \rvert \leq \frac{TkG}{n-k}.
\end{equation}
From here, we can use Markov's inequality and the upper bounds on $\bE\lvert \cY \rvert$ and $T$ to lower bound $\bP(\lvert \cY \rvert \leq k-1)$ as follows:
\begin{align}
    \label{eq:calY_prob_lb_1} \bP(\lvert \cY \rvert \leq k-1) &= 1 - \bP(\lvert \cY \rvert \geq k) \\
    \label{eq:calY_prob_lb_2} &\geq 1 - \frac{\bE\lvert\cY\rvert}{k} \\
    \label{eq:calY_prob_lb_3} &\geq 1 - \frac{TG}{n-k} \\
    \label{eq:calY_prob_lb_4} &\geq 2\delta.
\end{align}
Therefore, there is at least one defective item that is not recovered (and hence masked) with probability at least $2\delta$ if $T \leq (1-2\delta)\frac{n-k}{G}$.

Now we turn to characterizing the posterior distribution of $\cK$ given $(\boldsymbol{\cI}, \bfY)$. Using Bayes' theorem, we can write $\bP(\cK \mid \boldsymbol{\cI}, \bfY)$ as
\begin{align}
    \label{eq:bayes_1} \bP(\cK \mid \boldsymbol{\cI}, \bfY) &= \frac{\bP(\cK) \bP(\boldsymbol{\cI}, \bfY \mid \cK)}{\sum_{\cK' : \lvert \cK'\rvert = k}\bP(\cK') \bP(\boldsymbol{\cI}, \bfY \mid \cK')} \\
    \label{eq:bayes_2} &= \frac{\bP(\boldsymbol{\cI}, \bfY \mid \cK)}{\sum_{\cK' : \lvert \cK'\rvert = k}\bP(\boldsymbol{\cI}, \bfY \mid \cK')} \\
    \label{eq:bayes_3} &= \frac{\bP(\boldsymbol{\cI} \mid \cK) \bP(\bfY \mid \boldsymbol{\cI}, \cK)}{\sum_{\cK' : \lvert \cK'\rvert = k}\bP(\boldsymbol{\cI} \mid \cK) \bP(\bfY \mid \boldsymbol{\cI}, \cK')} \\
    \label{eq:bayes_4} &= \frac{ \bP(\bfY \mid \boldsymbol{\cI}, \cK)}{\sum_{\cK' : \lvert \cK'\rvert = k}\bP(\bfY \mid \boldsymbol{\cI}, \cK')}
\end{align}
where \eqref{eq:bayes_2} follows since $\cK$ is uniform over all ${n \choose k}$ possible outcomes, and \eqref{eq:bayes_4} follows since $\boldsymbol{\cI}$ is chosen non-adaptively (and is thus independent of $\cK$). Now, observe that since the test outcomes are noiseless, $\bfY$ is a deterministic function of $(\boldsymbol{\cI}, \cK)$. Hence, $\bP(\bfY \mid \boldsymbol{\cI}, \cK) \in \{0,1\}$, with it being characterized by
\begin{equation}
    \label{eq:IY_given_K} \bP( \bfY \mid\boldsymbol{\cI}, \cK) = \mathds{1}\left\{\cY \subseteq \cK \wedge \nexists \; i \in \cK, \; t \in [T] : i \lt Y_t \right\}.
\end{equation}
Put simply, this means that the observed test outcome vector $\bfY$ could have been generated by $(\boldsymbol{\cI}, \cK)$ if and only if $\cY \subseteq \cK$ (since otherwise there is a defective item not in $\cK$), and there are no tests $t$ such that the test outcome $Y_t$ appears before some $i \in \cK$ in the test (since otherwise there is a non-defective item in $\cK$). If $\cK$ meets both of the conditions described above, we call $\cK$ \emph{satisfying} (note that $\lvert \cZ \rvert \geq 1$ since the true defective set is always satisfying). Denoting the collection of satisfying sets as $\cZ$, it therefore follows that
\begin{equation}
    \label{eq:final_posterior_prob} \bP(\cK \mid \boldsymbol{\cI}, \bfY) = \frac{\mathds{1}\{\cK \in \cZ\}}{\lvert \cZ \rvert},
\end{equation}
which in turn implies that $(\cK \mid \boldsymbol{\cI}, \bfY) \sim {\rm Unif}(\cZ)$. Therefore, the error probability of the MAP decoder $\hat{\cK}_{{\rm MAP}}$ is $\bP(\hat{\cK}_{{\rm MAP}} \neq \cK \mid \boldsymbol{\cI}, \bfY) = 1- \frac{1}{\lvert \cZ \rvert}$, which also serves a lower bound on the (posterior) error probability for \emph{any} decoder $\hat{\cK}$.

To obtain a lower bound on the error probability of the MAP decoder (and thus any decoder), it remains to derive a (conditional) lower bound on $\lvert \cZ \rvert$. To do so, we introduce $\cM_D$ and $\cM_N$ as the events that there is at least one masked defective and one masked non-defective respectively. Since $\cM_D$ holds with at least probability $2\delta$ due to \eqref{eq:calY_prob_lb_1}-\eqref{eq:calY_prob_lb_4}, and $\cM_N$ holds deterministically and independently of $\cM_D$, it follows that $\bP(\cM_D \cap \cM_N) \geq 2\delta$. When the event $\cM_D \cap \cM_N$ occurs, we claim that this implies $\lvert \cZ \rvert \geq 2$. To see this, fix a masked defective item $i^* \in \cK$ and a masked non-defective item $j^* \notin \cK$ (whose existences are guaranteed by $\cM_D \cap \cM_N$), and consider the two previously discussed conditions for a set to be satisfying. As mentioned earlier, the true defective set $\cK$ is always a satisfying set, since $\bfY$ was generated by $(\boldsymbol{\cI}, \cK)$. Additionally, consider the set $\widetilde{\cK} \defeq (\cK \setminus \{i^*\}) \cup \{j^*\}$ (i.e., the set formed when swapping out $i^*$ with $j^*$). Since $i^*$ is masked, it is not the first defective item in any test, meaning that $i^* \notin \cY$. Additionally, the definition of $\widetilde{\cK}$ implies that $\cK \cap \widetilde{\cK} = \cK \setminus \{i^*\}$. Combining these observations then implies that $\cY \subseteq \cK \cap \widetilde{\cK} \subseteq \widetilde{\cK}$, which gives the first condition for $\widetilde{\cK}$ to be satisfying. The second condition is also satisfied by $\widetilde{\cK}$, since the items in $\widetilde{\cK} \setminus \{j^*\}$ are all defective, and since the definition of masking directly implies that $j^* \nless_t Y_t$ for all $t \in [T]$. Therefore $\widetilde{\cK}$ is also satisfying, which implies that $\lvert \cZ \rvert \geq 2$ when conditioned on $\cM_D \cap \cM_N$.

Writing $\cM \defeq \cM_D \cap \cM_N$, the above calculations and the law of total probability then imply that
\begin{align}
    \bP(\hat{\cK}_{\rm MAP} \neq \cK \mid \boldsymbol{\cI}, \bfY) &= \bP(\cM)\bP(\hat{\cK}_{\rm MAP} \neq \cK \mid \boldsymbol{\cI}, \bfY, \cM) \nonumber \\
    \label{eq:map_lb_1}  &\hspace{-8pt}+ \bP(\cM^c)\bP(\hat{\cK}_{\rm MAP} \neq \cK \mid \boldsymbol{\cI}, \bfY, \cM^c) \\
    &\geq 2\delta \left(1 - \frac{1}{2}\right) \nonumber \\
    \label{eq:map_lb_2} &\hspace{-8pt}+ \bP(\cM^c)\bP(\hat{\cK}_{\rm MAP} \neq \cK \mid \boldsymbol{\cI}, \bfY, \cM^c) \\
    \label{eq:mab_lb_3} &\geq \delta.
\end{align}
Since the error probability of the MAP decoder serves as a lower bound for error probability of \emph{any} decoder $\hat{\cK}$, it therefore follows that $\bP(\hat{\cK} \neq \cK \mid \boldsymbol{\cI}, \bfY) \geq \delta$. Since the above analysis holds for any fixed realization of $(\boldsymbol{\cI}, \bfY)$, taking an average over $(\boldsymbol{\cI}, \bfY)$ implies that $\bP(\hat{\cK} \neq \cK) \geq \delta$ if $T \leq (1-2\delta)\frac{n-k}{G}$. Taking the contrapositive of this statement gives the desired result. 

\section{Conclusion}
We have studied the cascaded group testing problem under the small error criterion, and provided achievability results for various different recovery criteria and levels of adaptivity in both the cases where test sizes are unconstrained, and upper bounded by some fixed quantity. In the constrained test size setting, we also provided a lower bound showing our result is optimal up to logarithmic factors. Perhaps the most immediate direction of future work is to establish whether order-wise matching upper and lower bounds can be obtained in both of these settings.
\bibliography{references}
\bibliographystyle{ieeetr}
\end{document}